\documentclass[10pt,journal,compsoc]{IEEEtran}
\usepackage[T1]{fontenc} %

\ifCLASSOPTIONcompsoc
  \usepackage[nocompress]{cite}
\else
  \usepackage{cite}
\fi

\usepackage{amsmath,amssymb,amsfonts}
\usepackage{algorithmic}
\usepackage{graphicx}
\usepackage{textcomp}
\usepackage[dvipsnames]{xcolor}
\usepackage[all]{nowidow}
\usepackage[caption=false]{subfig}
\usepackage{balance}
\usepackage[hidelinks]{hyperref}
\usepackage[capitalise,noabbrev,nameinlink]{cleveref}
\usepackage{tikz}  %
\usepackage{float}
\usepackage[all]{nowidow}
\usepackage{nicefrac}

\usepackage[disable]{todonotes}
\newcommand{\name}{\text{\textsc{Fusionize}}}

\newcommand{\eid}[2]{\MakeUppercase{#1-#2}}  %
\newcommand{\sid}[1]{\textit{setup\textsubscript{#1}}} %
\newcommand{\Sid}[1]{\textit{Setup\textsubscript{#1}}} %

\newcommand{\md}{\$pmi}

\newcommand{\fancytt}[1]{\texttt{%
        \hyphenchar\font=`\- %
        \hyphenpenalty=10000 %
        \exhyphenpenalty=-50 %
        #1\ignorespaces
    }}

\begin{document}
\title{\name{}++: Improving Serverless Application Performance Using Dynamic Task Inlining and Infrastructure Optimization}

\author{Trever~Schirmer,
    Joel~Scheuner,
    Tobias~Pfandzelter
    and~David~Bermbach%
    \IEEEcompsocitemizethanks{
        \IEEEcompsocthanksitem T.~Schirmer, T.~Pfandzelter, and D.~Bermbach are with the Scalable Software Systems research group at TU Berlin \& Einstein Center Digital Future, Berlin, Germany.\protect\\
        E-mails: \{ts, tp, db\}@3s.tu-berlin.de
        \IEEEcompsocthanksitem J.~Scheuner is with LocalStack, Switzerland.\protect\\
        E-mail: joel.scheuner@localstack.cloud
        \IEEEcompsocthanksitem This is the author copy of an article that has been accepted for publication in IEEE Transactions on Cloud Computing with DOI 10.1109/TCC.2024.3451108 (\url{https://doi.org/10.1109/TCC.2024.3451108}).}
        
}

\IEEEtitleabstractindextext{%
    \begin{abstract}

        The Function-as-a-Service (FaaS) execution model increases developer productivity by removing operational concerns such as managing hardware or software runtimes.
        Developers, however, still need to partition their applications into FaaS functions, which is error-prone and complex:
        Encapsulating only the smallest logical unit of an application as a FaaS function maximizes flexibility and reusability.
        Yet, it also leads to invocation overheads, additional cold starts, and may increase cost due to double billing during synchronous invocations.
        Conversely, deploying an entire application as a single FaaS function avoids these overheads but decreases flexibility.

        In this paper we present \name{}, a framework that automates optimizing for this trade-off by automatically fusing application code into an optimized multi-function composition.
        Developers only need to write fine-grained application code following the serverless model, while \name{} automatically fuses different parts of the application into FaaS functions, manages their interactions, and configures the underlying infrastructure.
        At runtime, it monitors application performance and adapts it to minimize request-response latency and costs.
        Real-world use cases show that \name{} can improve the deployment artifacts of the application, reducing both median request-response latency and cost of an example IoT application by more than 35\%. %
    \end{abstract}

    }

\maketitle

\IEEEraisesectionheading{\section{Introduction}
    \label{sec:introduction}}

\IEEEPARstart{W}{ith} the advent of serverless cloud computing, the Function-as-a-Service (FaaS) execution model has become a popular paradigm for large applications~\cite{Hendrickson2016-pw,paper_bermbach_cloud_engineering}.
In FaaS, developers write stateless, event-driven tasks that invoke each other to implement complex workflows~\cite{paper_grambow_befaas,jia2021nightcore,qi2022spright}.
Such tasks are deployed as FaaS functions on a cloud FaaS platform that abstracts operational concerns such as managing hardware or software runtimes, offering a flexible pay-as-you-go billing model\cite{paper_bermbach_cloud_engineering}.
Today, FaaS platforms are offered by all leading cloud providers, e.g., AWS Lambda\footnote{\url{https://aws.amazon.com/lambda}}, Google Cloud Functions\footnote{\url{https://cloud.google.com/functions}}, and Microsoft Azure Functions\footnote{\url{https://azure.microsoft.com/products/functions/}}, and are an area of major research interest~\cite{Akhtar_2020,Elgamal_2018,Czentye_2019,paper_bermbach_faas_coldstarts,Baldini_2017_Trilemma,Cordingly_2022_memorysizes,schirmer_2023_nightShift,paper_pfandzelter_tinyfaas}.

Despite the operational benefits of building applications as compositions of FaaS functions, we observe a gap between the developer-side logical view of complex applications and the performance and cost-efficiency characteristics of commercial cloud FaaS platforms.
On the one hand, application developers want to split their applications into isolated, single-purpose tasks that can be independently updated and worked on, may be dynamically recomposed, and that improve code reusability~\cite{Eismann_2021_Why}.
On the other hand, FaaS platforms incentivize large, monolithic FaaS functions as this avoids call overheads, cascading cold starts~\cite{paper_bermbach_faas_coldstarts, Mahgoub_2022_Wisefuse}, and double billing costs.
Furthermore, developers have to fine-tune the configuration parameters of the FaaS platform for every deployed function~\cite{Cordingly_2022_memorysizes,Kuhlenkamp_2022_configuration} -- to reduce effort, this also incentivizes developers to choose larger FaaS functions.%

\begin{figure}
    \centering
    \includegraphics[width=\linewidth]{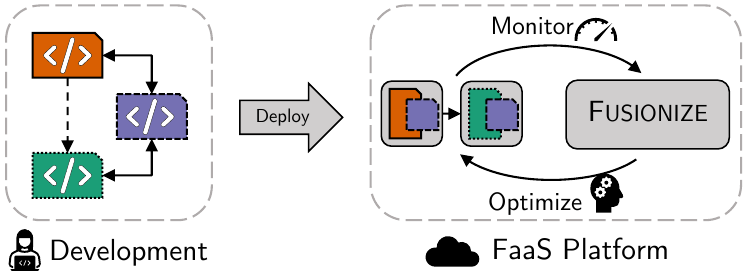}
    \caption{
        \name{} takes existing, unchanged FaaS tasks and optimizes their performance and cost-efficiency by iteratively modifying their deployment configuration and inlining tasks based on monitored performance. The developer-written tasks are deployed inside multiple FaaS functions, where they can be inlined instead of called remotely.
    }
    \label{fig:fusionize_high}
\end{figure}

In this paper, we address this gap with \name{}, a feedback-driven system for the automated configuration of composite task-based applications on cloud FaaS platforms.
A high-level overview of \name{} is shown in \cref{fig:fusionize_high}.
We borrow the concept of \emph{inlining} in compilers to expand remote FaaS functions calls with task source code where beneficial, a concept called \emph{function fusion}~\cite{Baldini_2017_Trilemma,Scheuner_2019}.
Function fusion can eliminate remote call overheads and restrict cold start cascades.
Further, we optimize the infrastructure configuration of deployed functions, such as allocated memory and CPU shares.
Notably, \name{} does not require additional configuration or application descriptions by software developers:
Without changes to the familiar FaaS programming model, \name{} takes existing FaaS applications and infers their call patterns, cost efficiency, and performance from their live execution behavior.
Then, \name{} iteratively optimizes the application deployment for cost efficiency and end-to-end performance.
If the behavior of the application changes, e.g., with changing load or with updates to the application, \name{} can automatically adapt to the new environment and optimize further.

We make the following main contributions in this paper, which is an extended version of the paper and system presented in~\cite{Schirmer_2022_fusionizePaper}:

\begin{itemize}
    \item We describe the main incongruities between the logical view of applications as a composition of event-driven tasks and optimal deployments on current cloud FaaS platforms (\cref{sec:issues}).
    \item We introduce \name{}, an automated approach for bridging the gap between development and deployment for FaaS applications (\cref{sec:system}).
    \item We present heuristics for the iterative optimization of FaaS applications in \cref{sec:heuristics}, creating a baseline and framework for future research on optimization approaches.
    \item We implement \name{} as an open-source prototype and present the results of extensive experimentation with, among others, two real world use cases on public cloud FaaS infrastructure (\cref{sec:evaluation}).%
    \item We discuss the limitations of our approach and derive avenues for future work (\cref{sec:discussion}).
\end{itemize}

\section{Challenges in FaaS Deployment}
\label{sec:issues}

The FaaS paradigm has introduced a new way of thinking about scalable, flexible applications as a composition of serverless functions that are invoked in an event-driven manner over the network and can call each other to implement complex workflows and business logic.
Unfortunately, the technology and pricing models of commercial FaaS platforms lag behind, with current platforms being optimized for applications encompassing only a single function.
A one-to-one mapping of the fine-grained software components provided by developers (in the following: \emph{tasks}) and the executable artifacts (in the following: \emph{functions}) that are deployed and instantiated in the cloud often yields suboptimal performance and cost efficiency.

\subsection*{Double Spending}

\begin{figure}
    \centering
    \includegraphics[width=\linewidth]{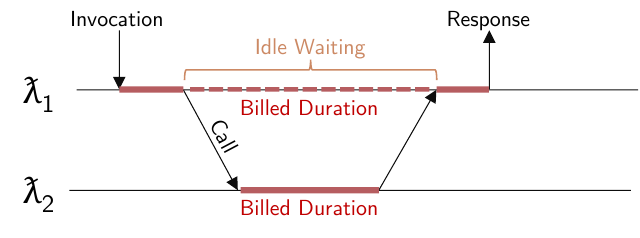}
    \caption{
        Synchronous invocations of functions can lead to double billing in the duration-based billing model of FaaS: While ${\lambda}_{1}$ calls ${\lambda}_{2}$, both functions incur costs.
    }
    \label{fig:doublespending}
\end{figure}

When a FaaS function makes a synchronous call to another function, i.e., it waits for the results of that call, the execution duration of the application is billed twice:
As shown in \cref{fig:doublespending}, the called function incurs costs, yet the waiting function is also billed, despite not performing any useful work.
Although the pattern of synchronous invocations between functions is crucial to build large, complex applications, the issue of double spending can be prohibitive from a cost perspective~\cite{Baldini_2017_Trilemma}.

\subsection*{Cascading Cold Starts}

\begin{figure}
    \centering
    \includegraphics[width=0.8\linewidth]{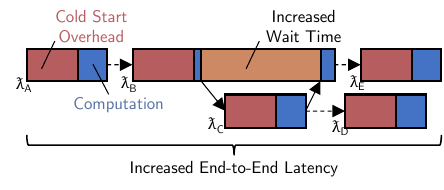}
    \caption{
        Cold start cascades occur when a series of function instances is executed for the first time, e.g., when load changes. The cold start overhead incurred for each execution increases the overall end-to-end latency and wait times in synchronous executions~\cite{paper_bermbach_faas_coldstarts}.
    }
    \label{fig:cascadingcoldstarts}
\end{figure}

The term ``cold start'' in FaaS refers to the overhead of creating a new function instance on-demand, i.e., when load changes and any existing instances are already occupied~\cite{paper_bermbach_faas_coldstarts}.
As illustrated in \cref{fig:cascadingcoldstarts}, in FaaS compositions with chains of functions executing in sequence, cold start overheads will accumulate, as every function instance of the chain is instantiated only when it is called~\cite{paper_bermbach_faas_coldstarts,daw2020xanadu}.
This increases both the application's end-to-end latency and its execution cost.

\subsection*{Infrastructure Configuration}

With current FaaS providers, developers have to manually choose the infrastructure configuration they wish to provision per function instance.
Most platforms allow users to specify the amount of memory a function has access to and provision CPU and I/O resources proportionally.
Adding more resources to a function can thus actually make its execution less expensive, as a lower execution duration can also reflect in lower cost per invocation with pay-per-ms FaaS pricing~\cite{Eismann_2021_Sizeless,Akhtar_2020,Cordingly_2022_memorysizes,Kuhlenkamp_2022_configuration}.
This involved configuration optimization partially negates the advantages of FaaS infrastructure abstraction.

\subsection*{Remote Function Call Overhead}

FaaS platforms are built for function invocation over the network, mostly using HTTP requests.
For external clients, accessing functions via an HTTP router makes sense, but it creates considerable invocation overhead for internal function calls~\cite{jia2021nightcore,qi2022spright}.
Grambow et al.~\cite{paper_grambow_befaas} find that the network transmission time, i.e., the overhead of calling a function from a function, in commercial FaaS platforms is on the order of 50ms.
This can be a significant overhead, especially for sequences of short-running functions.
With the pay-per-ms pricing model of FaaS platforms, this results not only in higher end-to-end latency for application but also increases execution costs.

\begin{figure*}
    \includegraphics[width=\textwidth]{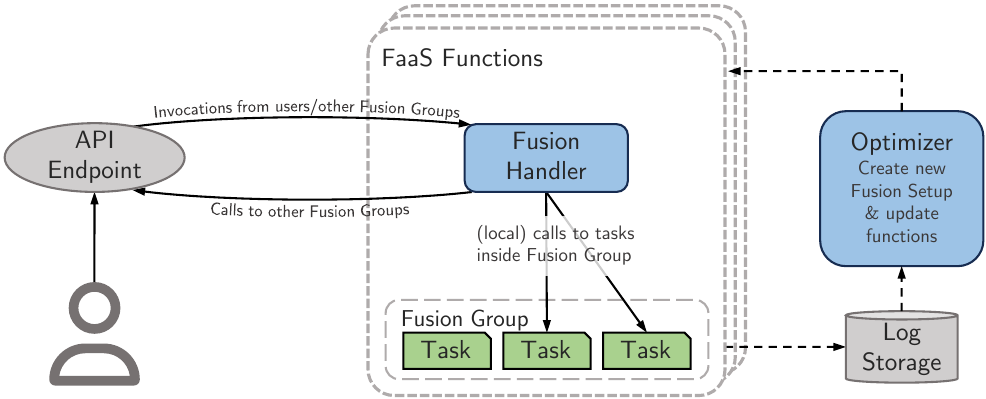}
    \caption{Overview of the Architecture of \name{}: Within a FaaS function, the fusion handler is responsible for interactions between tasks. Tasks can call other tasks, which are inlined if the task is in the same fusion group, or remotely handed off to the function responsible for another fusion group. The Optimizer regularly analyzes function logs to update the fusion setups.}
    \label{fig:approach:architecture}
\end{figure*}

\section{Automatic Function Fusion at Runtime}
\label{sec:system}

In this section, we give an overview of \name{} and describe how it can automatically adapt FaaS applications at runtime.
We start by describing key terms and definitions (\cref{subsec:terms}) before discussing \name{} and its components (\cref{subsec:approach}) and the resulting programming model (\cref{subsec:progmodel}).

\subsection{Terms and Definitions}
\label{subsec:terms}
As already mentioned above, we use the term \textbf{\emph{task}} to refer to the software function written by the developer and the term \textbf{\emph{function}} to refer to the executable deployment artifact (cf.~\cref*{fig:fusionize_high}: the left side shows tasks, the right side shows two functions containing tasks).
Each function contains a single \textbf{\emph{fusion group}}, i.e., a set of tasks that are executed as part of that specific function.
Calling a task that is in the same fusion group can be compared to inlining in compilers: the code for the other task is executed directly inside the same function.
When a task in another fusion group is called, execution is handed off to the function responsible for that group.
Tasks can be part of multiple fusion groups, i.e., we replicate some tasks to reduce the number of remote calls -- essentially, whenever the benefits of doing so outweigh the costs of replicating the task.
Please note that the FaaS provider will often also replicate functions by instantiating a deployed function more than once, possibly distributed over several physical machines.

To deploy the applications, we need to know all fusion groups, the infrastructure configuration of the functions, and where to hand off calls to other fusion groups.
We refer to this information as the \textbf{\emph{fusion setup}}.
At runtime, the fusion setup is changed dynamically to optimize the deployment artifacts.

In this paper, we use a short notation to describe fusion groups:
Tasks that are in a fusion group are put in parentheses, separated with commas.
Different fusion groups are separated with a hyphen.
For example, a developer might have specified three tasks \texttt{A}, \texttt{B}, and \texttt{C}, where \texttt{A} calls both \texttt{B} and \texttt{C}.
With the fusion groups \texttt{(A,B)-(C)}, the tasks \texttt{A} and \texttt{B} are in the same fusion group, and task \texttt{C} is in its own group.
This means that \texttt{A} calls \texttt{B} locally by inlining the code for task \texttt{B}, and calls from \texttt{A} to \texttt{C} are handed off to another function.
This notation is simplified and not able to show all possible fusion groups, as fusion groups are directed graphs.
For example, task \texttt{A} might inline task \texttt{B}, but not vice versa.

\subsection{The \name{} Approach}
\label{subsec:approach}
\name{} is a feedback-driven, autonomous deployment framework that collects and uses FaaS monitoring data to iteratively optimize the fusion setup.
This optimal fusion setup depends on developer preferences (regarding cost and performance goals, which may be in conflict) and is influenced by runtime effects.
\name{} is almost completely transparent to both developer and FaaS platform:
From the developer perspective, \name{} acts as a combination of monitoring and deployment tool (developers only need to adhere to the programming model described in \cref{subsec:progmodel}).
From the platform perspective, \name{} pretends to be a developer who tracks monitoring data and redeploys an application periodically.
Obviously, \name{} would ideally be part of the FaaS platform, but users of public clouds do not have any influence on that.
The focus of \name{} is on enabling an iterative optimization of the application as it behaves under real load, while keeping the additional load on developers low: as long as they adhere to the programming model (\cref{subsec:progmodel}), everything else is taken care of automatically without requiring the users to model anything or making any assumptions on how the application works.

Please note that \name{} does not do any modeling of application behavior, but instead only captures how the system behaves under real load.
For some scenarios, \name{} might, thus, not be as effective as some more modeling-heavy approaches that rely on test-runs~\cite{Mahgoub_2022_Wisefuse} or profiling data~\cite{Cordingly_2022_memorysizes}, as \name{} has to rely on accurate monitoring data.
In essence, \name{} can only consider calls that have actually been seen in practice.
This, however, can also be a strength in that \name{} ignores rare corner cases and, thus, can come to a better solution for the average application run.

As we show in a high-level overview of our approach in \cref{fig:approach:architecture}, \name{} has two main components: the \emph{Fusion Handler} and the \emph{Optimizer}.

\subsubsection*{Fusion Handler}

The Fusion Handler is responsible for dispatching requests between tasks.
As there is one function per fusion setup, the Fusion Handler is implemented as a distributed component co-deployed with every function.
From the FaaS provider perspective, the Fusion Handler is the endpoint for incoming calls to that function.
The Fusion Handler then forwards the call to the requested task locally.
If tasks call each other, the Fusion Handler logs all relevant data for this call to be used by the Optimizer.

\subsubsection*{Optimizer}

The Optimizer retrieves this monitoring data to derive the call graph of the application and annotate it with execution information, e.g., latency values.
In a second step, it uses an extensible optimization strategy module to derive and deploy an improved fusion setup.
We describe a first heuristic for this in \cref{sec:heuristics}.

The two extremes of function fusion, i.e., fusing nothing and fusing everything, both lead to suboptimal performance.
Fusing nothing leads to double spending, high remote call overhead and cascading cold starts, but the overall flexibility of the application is maximized and the infrastructure configuration can be optimized for individual tasks.
This leads to increased cost and latency for applications comprising short-running, synchronous tasks on the critical path.
These can be reduced by fusing all functions, which would however lead to suboptimal infrastructure configuration (e.g., if one task needs a lot of resources all other tasks also need to run in this more expensive instance) and might exhaust the resources of a function instance, thus, exceeding platform limits.
Additionally, if an application comprises a relatively small critical path but many uncritical resource-intensive tasks, the uncritical tasks would all need to be executed before the function can return its result to the caller.
For most applications, fusing some but not all tasks should improve performance.

We propose to use an adapted version of the continuous sampling plan CSP-1~\cite{dodge1943sampling,bermbach2011extendable} to decide when to run the Optimizer:
The algorithm, which originates from quality monitoring in a production line by sampling produced items, uses the quality of previous items to decide when to run the next quality inspection.
In the adapted version, we propose to compare cost and performance metrics of the monitoring snapshots considered during the previous and current Optimizer runs.
The larger the changes between runs, the sooner the Optimizer runs next.
This way, the Optimizer will run frequently for a newly deployed application but will still from time to time check performance and cost of ``older'' applications.
Applications can hence adapt to cost or performance changes resulting from external factors such as load profiles or changes to long-term platform performance\cite{Eismann_2022_stability,paper_bermbach_expect_the_unexpected}.
Within the Optimizer, the ``best'' fusion setup can be determined in various ways, e.g., optimizing for cost per invocation, request-response latency, or minimizing cold start impacts.
As part of the optimization strategy, application developers should here assign weights to different optimization goals.

\subsection{The \name{} Programming Model}
\label{subsec:progmodel}

From a FaaS developer perspective, the programming model is similar to standard FaaS programming.
The key difference is for calling other tasks:
Instead of calling remote FaaS functions directly, they tell the Fusion Handler to call the task, specifying whether the result is required synchronously or asynchronously.
All other operational tasks are handled by the Fusion Handler, which can transparently use different communication channels to communicate with other functions.
While we envision developers to write their code directly with function fusion in mind, this approach means that existing FaaS applications can easily be migrated to the \name{} programming model by detecting when a call to another function happens (i.e., via the SDK of the platform or by calling a specific endpoint) and rewriting that call into a call to the Fusion Handler.
Conceptually, all standard communication channels can easily be supported by the \name{} SDK.
Supporting multi-platform applications (e.g., Java and node.js) is a bit more challenging but is possible through technologies such as WebAssembly or the JVM, by using transpilation, or by deploying multi-platform functions which communicate via console-based process calls.

\section{Heuristically Optimizing FaaS Functions}
\label{sec:heuristics}

In this section, we propose a set of rules which allow \name{} to create a \emph{good} (but not necessarily the \emph{best}) fusion setup.
To achieve this, our heuristic analyzes multiple aspects of the application while it is running based on monitoring data.
Please note that the focus of this work is on the \textit{systems} aspects of \name{} and not on the optimization strategy, the further improvement of which we leave to future work.

\subsection*{Path Optimization}

Fusing tasks so they are executed in the same function as the calling task has the biggest impact on overall performance and cost (cf. \cref*{sec:evaluation}).
Consider the example of a setup in which one task calls another task and has to synchronously wait for the result before continuing.
If the called task is in another fusion group this will lead to double billing, likewise a remote call takes significantly longer than a local call.
Thus, those two tasks might benefit from being in one fusion group.

If a task is called asynchronously, remote execution of this task does not lead to double billing, as the calling task does not wait for called task to finish.
To free up the critical path, asynchronous tasks should thus be handed off to another function, i.e., put into another fusion group.
These general rules might not be optimal in every case: the overhead to make a call to another function (\textasciitilde{}50ms) might exceed the time needed for executing the task locally, in which case it might be better to run it locally.
This might even be true for longer running tasks during cold starts to prevent cascading cold starts~\cite{daw2020xanadu}.
But, again, please note that the focus of this paper is not the Optimizer but the fusion framework around it.
We expect future research to identify better fusion strategies than the heuristic presented here which aims for a good but not optimal solution.

\begin{figure}[t]
    \centering
    \includegraphics[width=0.9\columnwidth]{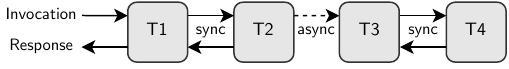}
    \caption{Example call graph that could benefit from function fusion: Task \texttt{T1} and \texttt{T2} need to finish before a response can be sent, and could thus be fused to reduce costs. \texttt{T3} is only called asynchronously, and should thus be moved to another fusion group.}
    \label{fig:approach:example}
\end{figure}

An example for path optimization is shown in \cref{fig:approach:example}: \texttt{T1} receives a request, makes a synchronous call to \texttt{T2}, which makes an asynchronous call to \texttt{T3} (e.g., for logging purposes) which finally calls \texttt{T4} synchronously.
After path optimization, all synchronous tasks are fused, and all asynchronous tasks are split off, leading to the fusion groups \texttt{(T1,T2)-(T3,T4)}.

\subsection*{Infrastructure Optimization}
Once the path has been optimized, changing the infrastructure configuration of the underlying function can optimize deployment goals even further.
While we were able to find a heuristic for path optimization, it is hard to predict the optimal infrastructure configuration without benchmarking all of them~\cite{Cordingly_2022_memorysizes,Kuhlenkamp_2022_configuration}.
While it is possible to measure some aspects of the behavior of a task (e.g., its memory and CPU usage), it is not possible to predict which one will be the dominant influence factor on total cost per execution.
Thus, we need to check every possible configuration of every group (but not of every task).
This can be accelerated if it is run after path optimization:
Since the functions all call each other asynchronously, they do not have wait for each other.
This way they can try out resource configurations in parallel without influencing each other.
The Optimizer can identify the optimal infrastructure configuration for every function after trying every memory size on it once.
This greatly reduces the number of optimization runs necessary compared to checking every possible path.

\subsection*{Combined Heuristic}
\begin{figure}[t]
    \centering
    \includegraphics[width=0.45\textwidth]{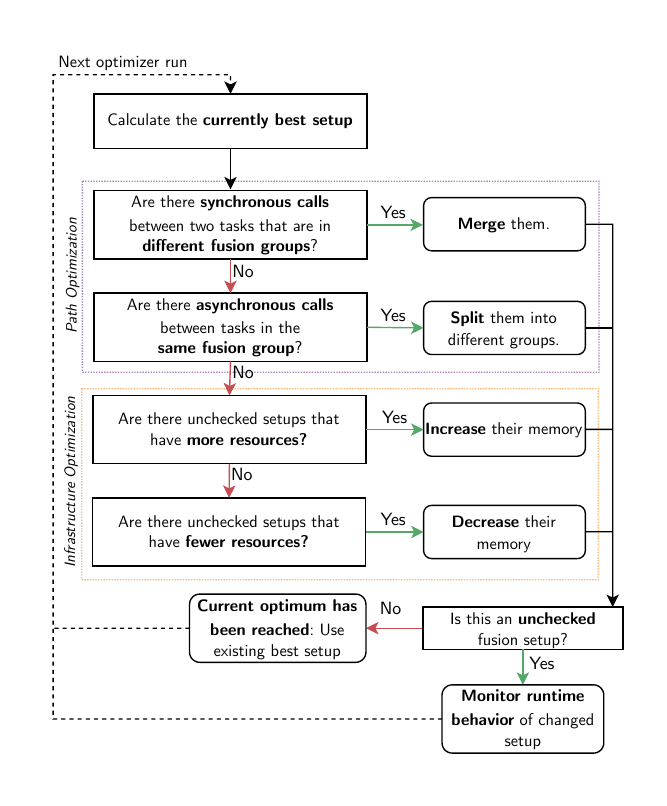}
    \caption{Our heuristic calculates the setup that currently performs best and tries to improve it in two phases: First it performs path optimization, then the optimal path is used to perform infrastructure optimization.
    }
    \label{fig:optimizationAlgo}
\end{figure}

Bringing all these aspects together, we propose the heuristic shown in \cref{fig:optimizationAlgo}.
During the path optimization phase, the optimizer moves all synchronous tasks to the same fusion group and all asynchronous tasks to remote fusion groups, repeatedly checking whether the change improved application performance.
Once the fastest path has been found, the groups are checked for the optimal infrastructure configuration.

There are further influence factors that could be considered here:
For example, inconsistencies in the performance of the underlying infrastructure or changes to remote services that are called by tasks might influence the optimal fusion setup and could be detected during execution~\cite{schirmer_2023_nightShift}.
As another example, overall cost might be lowest when only using specific infrastructure configurations.
The application performance might also be influenced by outside factors, such as changes in secondary services, code changes to the application, or long term changes to platform behavior.
By adjusting the weight of the monitoring data to favor recent measurements, the optimizer adapts to these changes by itself over time.
It is also possible to add these outside factors as inputs to the Optimizer itself so that they can directly be taken into account.
We leave these to future work as our focus is on the systems aspects of \name{} and not on fine-tuning Optimizer strategies.

\begin{figure*}
    \centering
    \includegraphics[width=\textwidth]{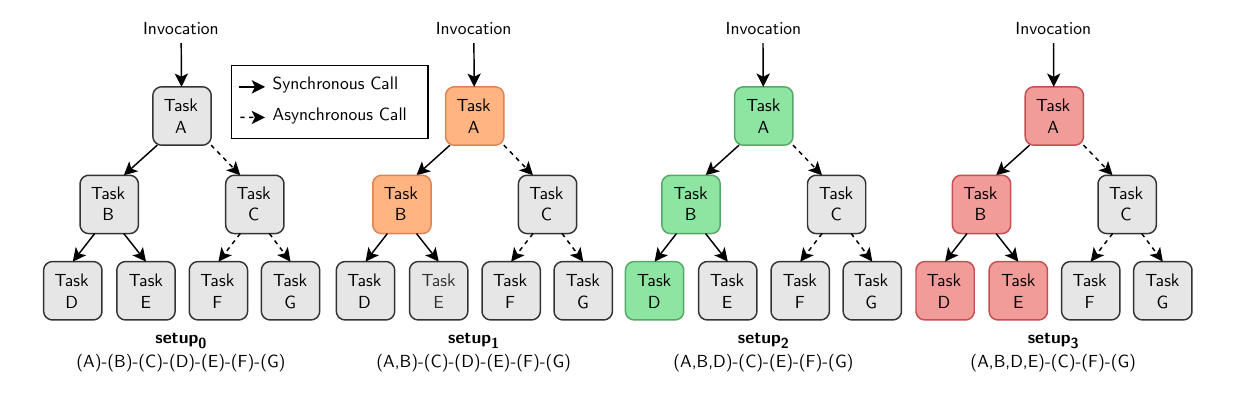}
    \caption{
        This figure shows the call graph of the TREE use case.
        Non-leaf tasks call two tasks each, with one side of the call tree containing synchronous tasks and the other side containing asynchronous tasks.
        From left to right, the different call graphs checked by the optimizer are shown.
        During path optimization, the optimizer starts with the initial setup \sid{base} and changes the fusion group of one task per step until it reaches \sid{3}, which is the path-optimized setup in which all synchronously called tasks have been fused.
        Afterwards, the optimizer performs infrastructure optimization on the path-optimized setup (\sid{4} to \sid{12}).
        For this figure, all colored tasks are fused into one fusion group, all other tasks are in their own fusion group.
    }
    \label{fig:eval:allInOne}
\end{figure*}

\section{Evaluation}
\label{sec:evaluation}

Our evaluation of \name{} entails a prototypical implementation (\cref{subsec:implementation}) and three use case applications implemented for our experiments (\cref{subsec:usecases}).
We describe experiment designs and parameters (\cref{subsec:experiments}), present the results of experimenting with all three applications (\cref{subsec:results}), and quantify overheads introduced through the use of \name{} (\cref{subsec:overhead}).
We make all artifacts and our prototype available as open-source software.\footnote{\url{https://github.com/umbrellerde/functionfusion}}

\subsection{Prototype Implementation}
\label{subsec:implementation}

We implement a prototype of \name{} for AWS Lambda.
We focus on Node.js for this proof-of-concept as it is the most widely-used runtime on AWS Lambda~\cite{Eismann_2021_Review} and an interpreted language, which simplifies dynamic loading of code.
We note, however, that our approach is extensible for other programming languages and FaaS platforms.

Application-internal task invocations use the embedded Fusion Handler that routes the call as a local JavaScript function call for fused tasks or externally over HTTPS for tasks that are deployed in a different fusion group.
The Fusion Handler logs monitoring data via the platform logging mechanisms, which then are read by the Optimizer via the platform-specific log extraction mechanism.
To explore different resource configurations, every function that handles a fusion group is deployed once for every configured memory size.

\subsection{Use Case Applications}
\label{subsec:usecases}

With our experiments, we aim to show the wide applicability of \name{} to different applications as well as the benefits which can be achieved through this.
For this reason, we implement three example applications (TREE, IOT, WEB):

\subsubsection{TREE: Synthetic Fan-Out Application}
The TREE use case is built as a synthetic fan-out application to fully demonstrate the features of \name{}.
The TREE application arranges a number of tasks as a call tree in which each task aside from the leaf nodes calls two other tasks (see also \cref{fig:eval:allInOne}).
One of the subtrees of the root task contains only synchronous calls to lightweight tasks which do not run complex computations.
The other subtree contains only asynchronous calls and all tasks are compute-intensive and perform mathematical operations in two threads.
As a result, compute-intensive tasks will benefit from multicore processing while the lightweight tasks should for cost-efficiency reasons be run with the smallest possible resource size.

\subsubsection{IOT: An Internet of Things Application}
The IOT use case aims to show the behavior of \name{} when used with a realistic Internet of Things application which depends on external services, a common use case for serverless computing~\cite{Eismann_2021_3c, Castro_2019, paper_grambow_befaas,paper_bermbach_cloud_engineering}.
In this use case, roadside sensors measure temperature, noise level, and air quality.
The sensor values are then analyzed by several tasks to identify anomalous conditions, e.g, for warning purposes.
All readings as well as necessary persistent task state are stored in a serverless database (AWS DynamoDB).
We show an overview of the call graph of this application in \cref{fig:eval:iot}.
As in the TREE use case, all asynchronously called tasks run CPU-intensive operations to simulate typical machine learning workloads.
Additionally, the tasks \texttt{AS}, \texttt{CSA}, \texttt{DJ}, and \texttt{SE} write data into the database, while \texttt{CSL} sends two read queries to the database before writing data itself.

\subsubsection{WEB: A Web Shop Application}
\begin{figure*}
    \centering
    \includegraphics[width=\textwidth]{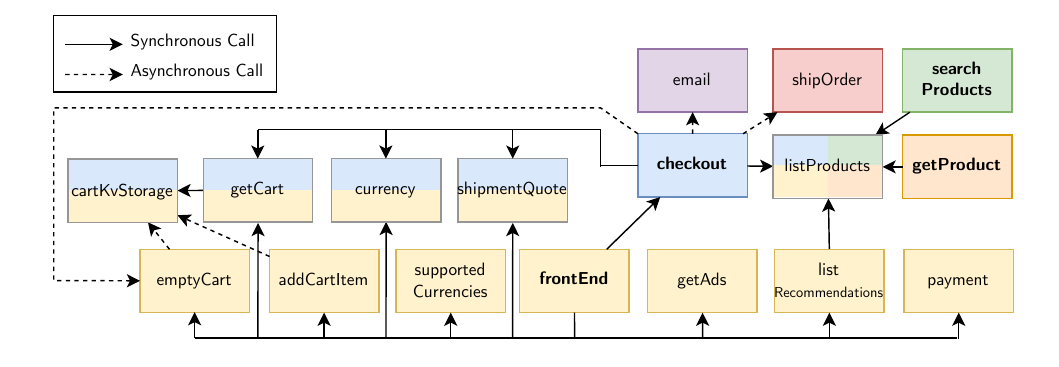}
    \caption{Call graph for the WEB use case with the fusion setups marked in different colors.
        Note that there are four tasks that are directly called by users (marked in bold in the figure) and that some tasks (e.g., \texttt{currency}) are fused into multiple functions.
        \texttt{listProducts} is called synchronously by four root tasks, so that it is fused into four different fusion groups.
    }
    \label{fig:eval:web}
\end{figure*}

The WEB use case aims to further evaluate how \name{} performs with complex applications, specifically how it deals with different call graphs.
For this, we adapted a web shop scenario from a microservice demo application~\cite{GCP_Microservices_Website}.
The application consists of 17 tasks shown in \cref{fig:eval:web}:
Customers can browse items, get recommended items, add items to their cart, and do a checkout where shipping costs are calculated, an e-mail is sent out, and a credit card transaction is performed.
These operations often write or read data from a serverless database (AWS DynamoDB).
There are different operations users can perform with this application: by viewing the frontend, users can see their current cart, recommended products, and the total shipping costs.
Users can, however, also call all these tasks directly without having to call the frontend task first.
With this, we can study how \name{} fares in the presence of alternative call graphs, where some users call the frontend while others call the relevant tasks directly.
In the previous two use cases, the same root task is used for all invocations.
In this use case, we perform a typical user flow where three tasks are called (adding a product to the cart, navigating to the front page, completing a checkout).

\begin{figure*}
    \centering
    \includegraphics[width=\textwidth]{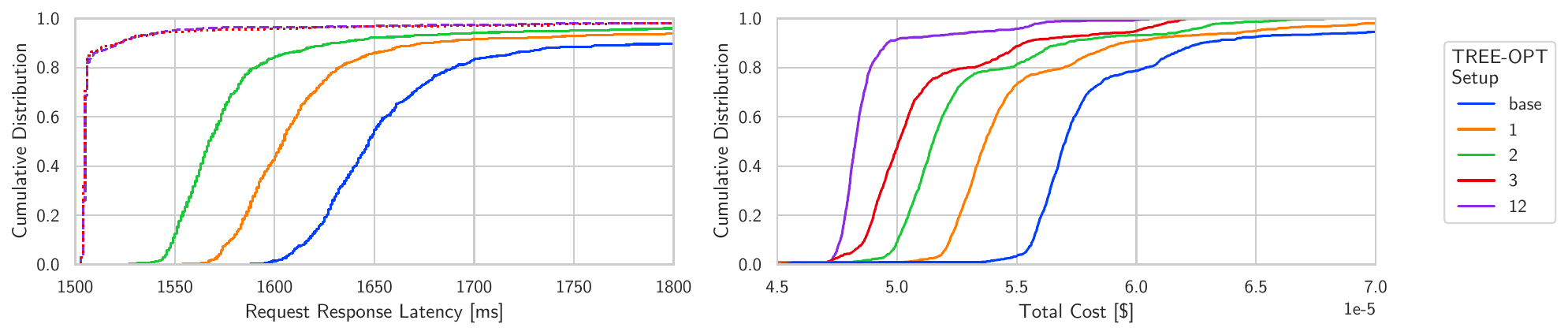}
    \caption{In the \eid{Tree}{Opt} experiment, \name{} iteratively improves fusion setup performance: The first two optimization runs decrease cost and $rr$. After path optimizations, the optimizer tries all configured function sizes using the fusion groups in \sid{3}. For readability reasons, we only show the final fusion setup \sid{12} where every function uses the cheapest memory size. As \sid{3} and \sid{12} use the same memory size for task A, they have a similar $rr$.
    }
    \label{fig:eval:split:full}
\end{figure*}

\begin{figure*}
    \centering
    \includegraphics[width=\textwidth]{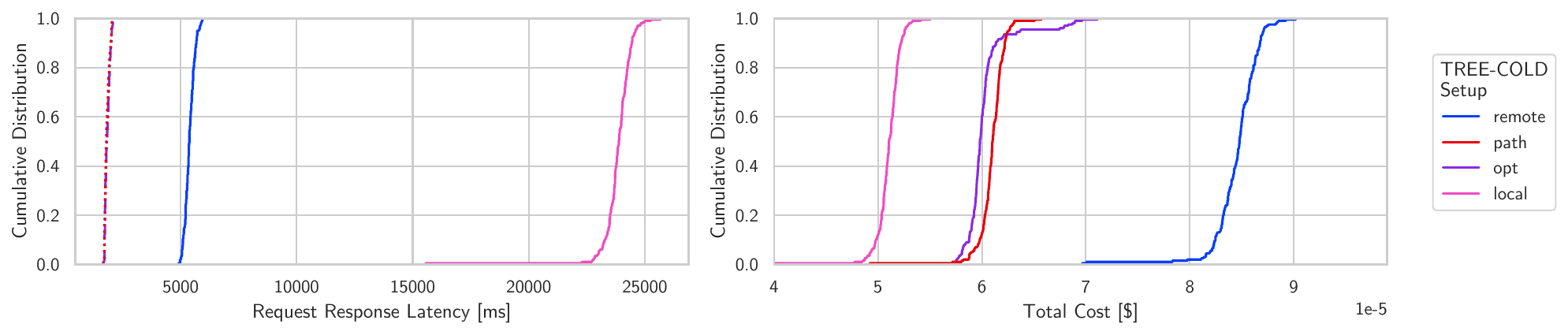}
    \caption{This graph shows the results of the \eid{Tree}{Cold} experiment. Both \sid{path} and \sid{opt} use the same memory size for the initial task A, resulting in a similar $rr$. \Sid{local} has the lowest cost for cold starts, but its median request-response latency is four times higher than \sid{opt}.
    }
    \label{fig:eval:split:cold}
\end{figure*}

\begin{figure*}
    \centering
    \includegraphics[width=\textwidth]{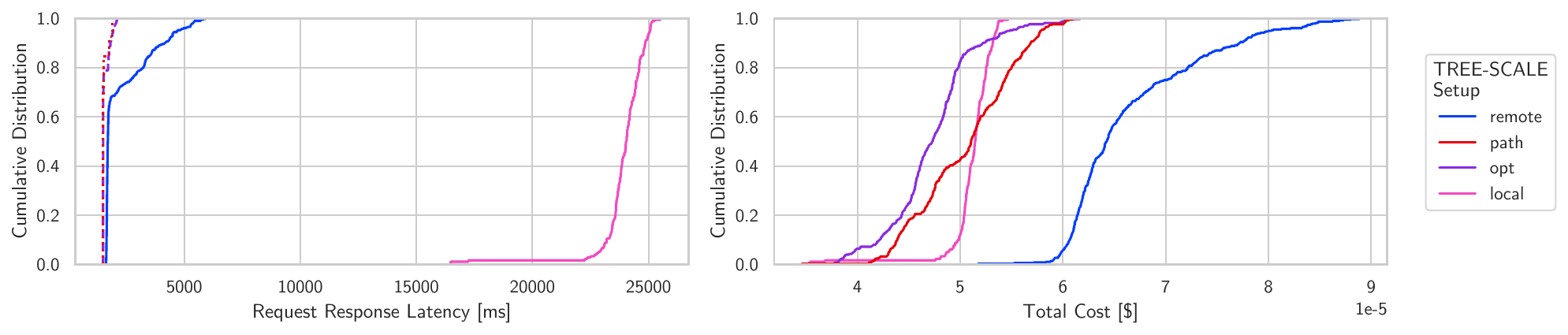}
    \caption{This graph shows results of the \eid{Tree}{Scale} experiment. While \sid{path} and \sid{opt} have a comparable $rr$, memory optimization has decreased total cost. While \sid{local} is only slightly more expensive than \sid{opt}, it has a higher $rr$.}
    \label{fig:eval:split:bursty}
\end{figure*}

\subsection{Experiment Design and Setup}
\label{subsec:experiments}
For each of the three use cases described above, we run three experiments (*-OPT, *-COLD, and *-SCALE) which we describe in the following.

\subsubsection{\eid{Tree}{Opt}, \eid{IoT}{Opt}, \eid{Web}{Opt}}
With these experiments, we show that \name{} can iteratively optimize the request-response latency ($rr$, i.e., the latency observed at a client calling the FaaS application) and total cost of invocations.
We put our prototype under a load of ten requests per second (rps) for 100 seconds.
For our experiments, we do not use CSP-1 (cf.~\cref{subsec:approach}) to decide on the best time to run the Optimizer.
Instead, we run it after every 1,000 requests until it has found the best fusion setup.
This allows us to collect the same number of requests for every step of the Optimizer in this evaluation, thus, easing our analysis and presentation of experiment results.
We then calculate median ($rr_{med}$) request-response latency and the average total cost ($cost$) of invocations.
In the initial fusion setup in all optimizer experiments, all tasks are in their own fusion group (\sid{base}).
This way, every call is to a remote function.
Without \name{}, functions would be deployed this way in a serverless system to maximize flexibility and reusability.
We then name setups in the order they are checked by the Optimizer (\sid{1}, \sid{2}, etc.)

\subsubsection{\eid{Tree}{Cold}, \eid{IoT}{Cold}, \eid{Web}{Cold}}
With these experiments, we study how frequently the different fusion setups found by \name{} encounter cold starts.
For this, we compare (i) the baseline setup in which all tasks are called remotely, thus, maximizing cascading cold starts (\sid{remote}), (ii)~the setup in which every task is called locally, thus minimizing the impact of cascading cold starts (\sid{local}), (iii)~the path-optimized fusion setup found by the optimizer in the respective *-OPT experiment (\sid{path}), and (iv)~the setup that has also been infrastructure-optimized (\sid{opt}).
To test specifically for function cold starts, we change an otherwise unneeded environment variable in every function, which leads to the platform shutting down any running instances.

\subsubsection{\eid{Tree}{Scale}, \eid{IoT}{Scale}, \eid{Web}{Scale}}
With these experiments, we evaluate how \name{} reacts to changing workloads by constantly increasing the load on the system.
The fusion setups used in these experiments are the same as in the respective *-COLD experiments.
Starting with 5rps, the load is increased by 5rps every two seconds up to 40rps (after 18 seconds), which leads to approximately 5 cold starts every two seconds.

For all experiments presented here, we configure all functions to start with 128MB of memory by default, but let the optimizer try the following memory sizes (in MB): 768, 1024, 1536, 1650, 2048, 3000, 4096, 6144.
AWS Lambda allocates CPU power proportionally to memory size, giving the functions access to between 0.08 and 3.5vCPUs~\cite{Cordingly_2022_memorysizes}.
The function with 1,650MB memory has access to around one vCPU.

\begin{figure*}
    \centering
    \includegraphics[width=\textwidth]{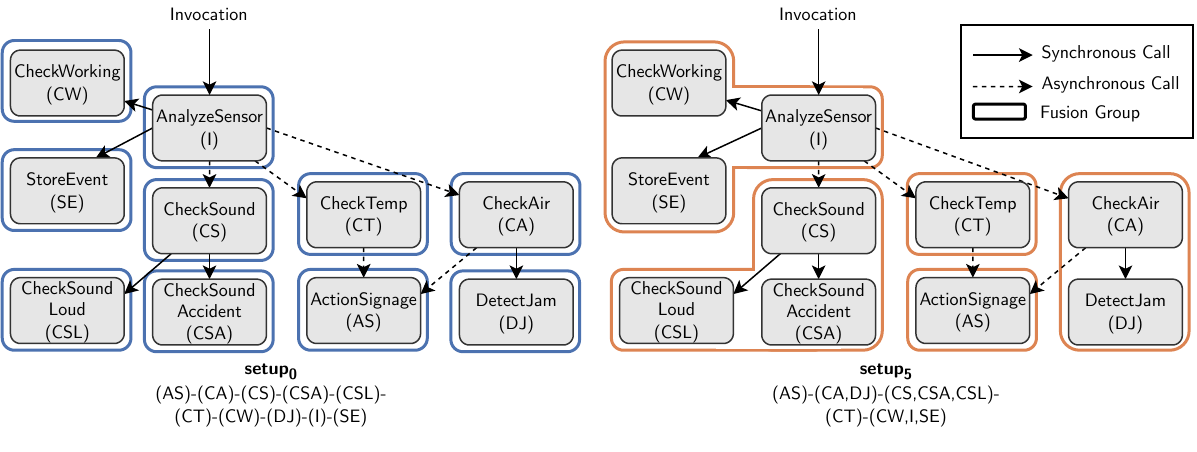}
    \caption{Call graph for the IoT application with the initial (\sid{base}) and path optimized (\sid{5}) fusion setup marked. In \sid{5}, all synchronous call chains are in the same fusion group. All setups after \sid{5} have the same call graph, but other infrastructure configurations.
    }
    \label{fig:eval:iot}
\end{figure*}

\subsection{Results}\label{subsec:results}
In this section, we present the results for all experiments ordered by use case -- first TREE, then IOT, then WEB.

\subsubsection*{\eid{Tree}{Opt}}
\label{sec:eval:split:full}
\label{subsec:tree}
The results of the full experiment are shown in \cref{fig:eval:split:full}.
In the call graph shown in \cref{fig:eval:allInOne}, we mark all fusion setups that were tried out by \name{} during our experiments.
The Optimizer first changes the initial setup \sid{base} to \sid{1} by fusing tasks \texttt{A} and \texttt{E} together.
In \sid{2}, task \texttt{D} is also put in this group.
For \sid{3}, task \texttt{B} is also added to this group leading to the fusion setup in which all synchronous calls are local \fancytt{(A,B,D,E)-(C)-(F)-(G)}.
This is the path-optimized setup (\sid{path}), where all synchronous calls are called locally, and all asynchronous calls are remote.
The Optimizer then tries out every remaining function in every configured memory size (except the initial memory size which has already been checked), until it reaches the final setup \sid{12}, in which every function is configured to run with the memory size that leads to the lowest total cost (\sid{opt}).
In this setup, the function handling the group \fancytt{(A,B,D,E)} has 128MB of RAM, \fancytt{(C)} has 1,024MB, and \fancytt{(F)} and \fancytt{(G)} are allocated 1,536MB each.
Applying the heuristic reduced $rr_{med}$ from 1.6s to 1.5s and reduced the average total cost per invocation by 18\% from 57.04\md{} to 48.26\md{}
(we use \md{}, i.e., USD per million invocations, as our monetary unit).
Note further that variability and tail latency of our application are also reduced through optimization.

\subsubsection*{\eid{Tree}{Cold}}
\label{subsec:tree:cold}

The results of the tree cold start experiments are shown in \cref{fig:eval:split:cold}.
Noticeably, calling every task locally during a cold start (\sid{local}) minimizes total cost for this invocation.
However, the request-response latency is substantially higher, since no tasks can be offloaded or called in parallel.
$rr_{med}$ of \sid{local} is 23,882ms, whereas \sid{remote} is more than four times as fast (5,380ms).
\Sid{path} and \sid{opt} are more than 12 times faster ($\sim$1,800ms) than \sid{remote}.
The median total cost of an invocation is 84\md{} for the remote setup.
The path optimized setup \sid{path} (61\md{}), the infrastructure optimized setup \sid{opt} (59.8\md{}), and the local setup \sid{local} (51\md{}) are all less expensive.

\subsubsection*{\eid{Tree}{Scale}}
\label{subsec:tree:scale}

The results of the \eid{Tree}{Opt} experiment are shown in \cref{fig:eval:split:bursty}.
While \sid{local}, \sid{path}, and \sid{opt} are less expensive than \sid{remote}, $rr$ is minimized by \sid{path} and \sid{opt}.
This shows that the optimized fusion setups also improve performance under different kinds of load.
The average cost is 64.2\md{} for \sid{remote}, 51\md{} for \sid{path} and \sid{local}, and 47.3\md{} for \sid{opt} (36\% reduction from \sid{remote}).

\subsubsection*{TREE: Discussion}

\name{} is able to reduce cost by 20-50\% compared to the baseline while also decreasing request response latency in all three experiments.
In \eid{Tree}{Cold}, \sid{local} is 17\% less expensive than \sid{opt}. However, \sid{local} is also 13 times slower than \sid{opt}.
This shows that \name{} can be used to improve performance and cost of serverless applications in different experiment setups.

\subsubsection*{\eid{IoT}{Opt}}

\begin{figure*}
    \centering
    \includegraphics[width=\textwidth]{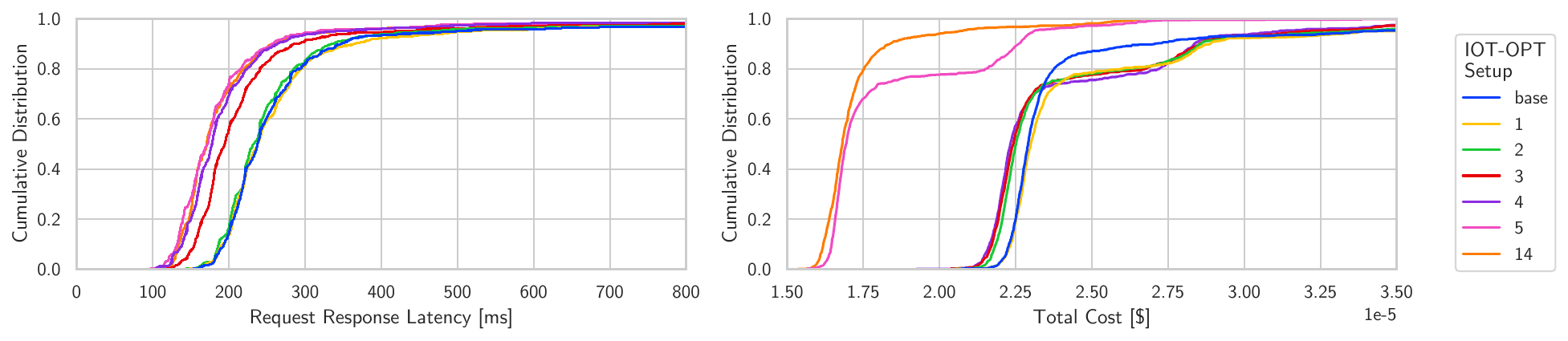}
    \caption{In \eid{IoT}{Opt}, $rr_{med}$ is reduced from 237ms to 171ms.
        While \sid{5} (path optimized) and \sid{14} (infrastructure optimized) have a very similar $rr$, the median invocation of \sid{14} is slightly cheaper.}
    \label{fig:eval:iot:full}
\end{figure*}

In the \eid{IoT}{Opt} experiment (cf.~\cref{fig:eval:iot:full}), the Optimizer tries four fusion setups before reaching the path optimized fusion setup \sid{5} (\fancytt{(AS)-(CA,DJ)-(CS,CSA,CSL)-(CT)-(CW,I,SE)}).
\Cref{fig:eval:iot} shows the overall call graph of TREE as well as \sid{base} and \sid{5} from this experiment.
After \sid{5}, the optimizer deploys and compares all eight possible other memory sizes, arriving at \sid{14} as the optimal setup  where total cost is minimized by using the smallest available function size of 128MB for all functions except for \texttt{(AS)}, which has a memory size of 1,650MB.
This is the case since most tasks read or write from DynamoDB and are thus I/O-bound, i.e., more function resources do not affect function latency.
The median total cost of the first four setups drops from 22.9\md{} in \sid{base} to 16.9\md{} in \sid{5}, while \sid{14} has a median total cost of 16.8\md{}, thus reducing cost by 36\%.
The median request-response latency $rr$ is reduced by 37\% from 237ms in \sid{base} to 171ms in \sid{14}.
\Sid{5} and \sid{14} have almost the same request-response latency since both fusion setups have the same call graph and execute the initial task \texttt{A} using the same infrastructure configuration (128MB).

\subsubsection*{\eid{IoT}{Cold}}
\begin{figure*}
    \centering
    \includegraphics[width=\textwidth]{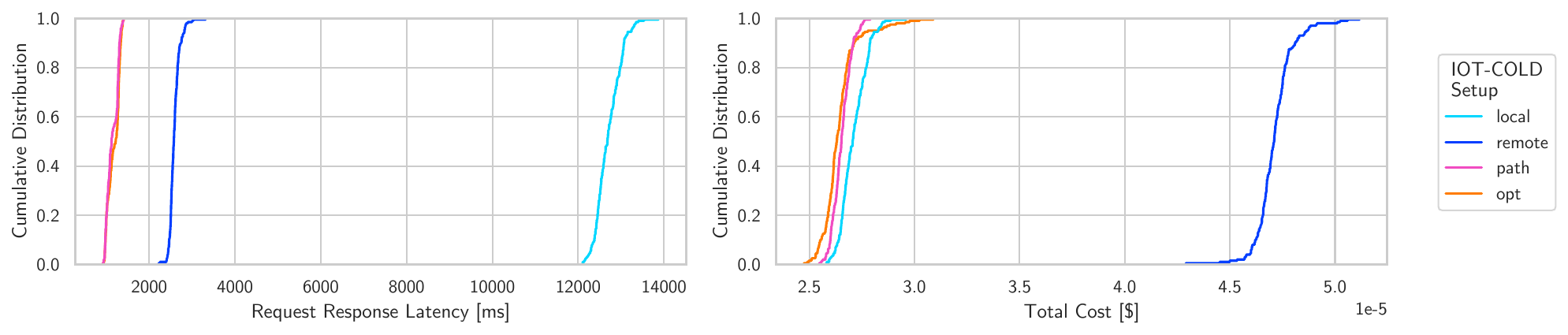}
    \caption{In the \eid{IoT}{Cold} experiment, \sid{opt} avoids cascading cold starts, leading to a 53\% (1,000ms) lower $rr_{med}$ than \sid{remote}. Calling all tasks locally has a comparable cost to the optimized setup, but has a far higher $rr$.
    }
    \label{fig:eval:iot:cold}
\end{figure*}

The results for the cold start experiment are shown in \cref{fig:eval:iot:cold}.
Overall, invocations for \sid{remote} have the highest cost, since this setup creates many cascading cold starts.
\Sid{local} is only slightly more expensive than \sid{path} and \sid{opt}, but its median request-response latency is more than ten times higher (12,681ms compared to 1,217ms).
The median cost for \sid{remote} is 47\md{}, and between 26 and 27\md{} for all other setups ($\geq$74\% reduction).

\subsubsection*{\eid{IoT}{Scale}}
\begin{figure*}
    \centering
    \includegraphics[width=\textwidth]{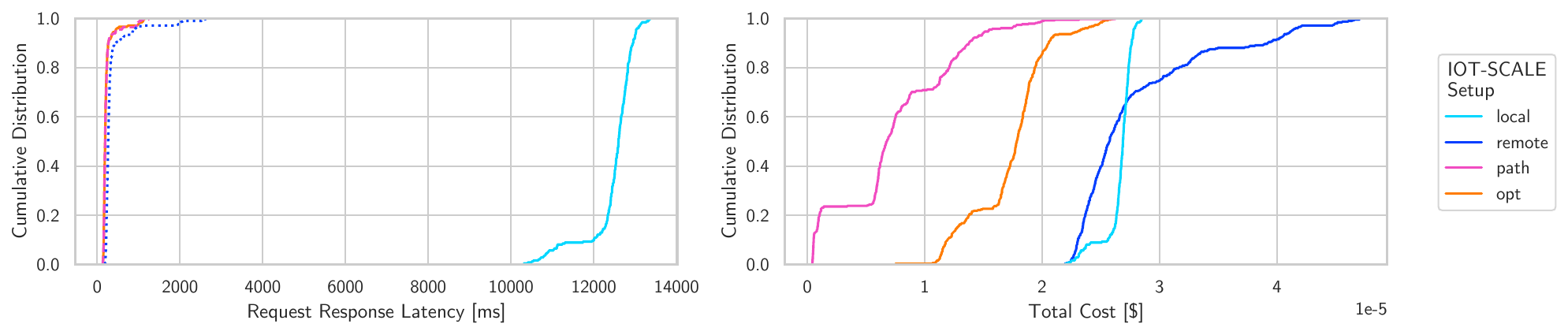}
    \caption{In the \eid{IoT}{Scale} experiment, the scaling workload leads to many cold starts.
        \Sid{remote} and \sid{local} have a similar median total cost of around 2.6\md{}, while \sid{opt} has an almost four times lower median cost of 0.67\md{}.}
    \label{fig:eval:iot:bursty}
\end{figure*}

The scalability experiment (cf.~\cref{fig:eval:iot:bursty}) shows comparable results to the cold start experiment \eid{Tree}{Cold}, except that \sid{optim} is now less expensive than \sid{local}.
In this case, cost would be minimized by always using \sid{opt}.
The median total cost for \sid{remote} is 25.7\md{}, \sid{local} is 26.8\md{}, \sid{path} is 6.75\md{}, and \sid{opt} is 17.8\md{}.
This makes the optimized setup 44\% less expensive than the default setup.
The median request-response latency is 12,597ms for \sid{local}, 262ms for \sid{remote}, and 190ms for \sid{path} and \sid{opt} (37\% faster than \sid{remote}).

\subsubsection*{IOT: Discussion}

Overall, these results show that \name{} can also be used to minimize cost and reduce request-response latency in a more complex use case with complicated call patterns and calls to remote services.
The setup found by the Optimizer is significantly less expensive and faster than the other setups in all three experiment setups.

\subsubsection*{\eid{Web}{Opt}}

\begin{figure*}
    \centering
    \includegraphics[width=\textwidth]{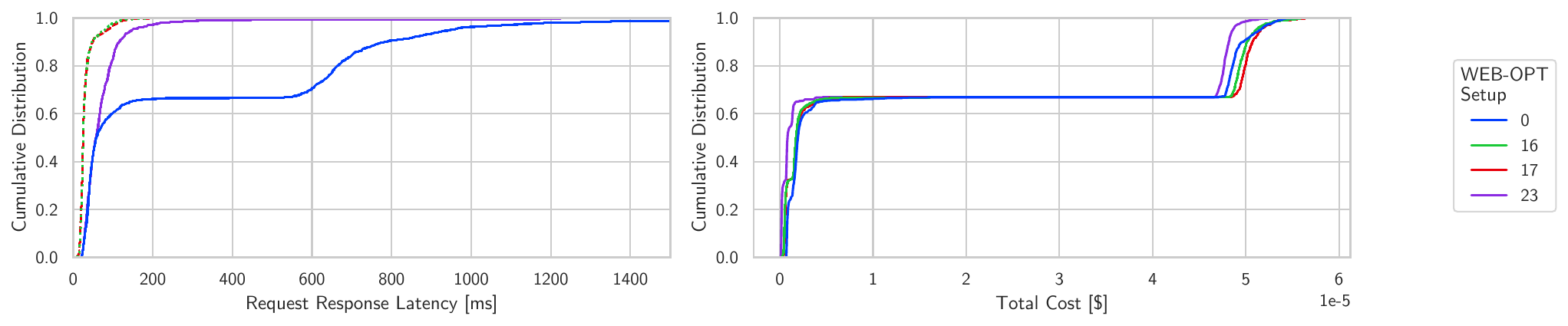}
    \caption{The web shop application comprises 17 different tasks with complicated call patterns. In the \eid{Web}{Opt} experiment, the optimizer tried twelve configurations (not shown here due to space constraints) before arriving at \sid{13}.
        Afterwards all memory sizes are checked, until the infrastructure optimized setup \sid{23} is found.
        This graph also shows \sid{16} and \sid{17}, where every function is configured with 1536MB and 1650MB RAM respectively.
        These are the two memory sizes where the function has access to slightly less and slightly more than one vCPU.
        In this experiment, \sid{23} has a lower median cost than \sid{16} and \sid{17}, but has a higher $rr_{med}$.
        Noticeably, \sid{13} and \sid{23} are the same fusion setup.
        This means that the infrastructure optimized setup always uses the smallest available infrastructure configuration.
        We only show \sid{23} in this graph, as it has almost complete overlap with \sid{13}.
    }
    \label{fig:eval:shop:full}
\end{figure*}

In the \eid{Web}{Opt} experiment (cf.~\cref{fig:eval:shop:full}), the optimizer tries twelve other fusion setups before arriving at the path optimized fusion setup \sid{13}, which again fuses all synchronous calls locally and puts all asynchronous calls into remote groups.
The optimizer then checks all other memory sizes and arrives at \sid{22}, where every function has the optimal memory size.
In contrast to the previous experiments, the infrastructure optimized setup is the same as the path-optimized setup (i.e., \sid{13} = \sid{22}), as cost is minimized when every function uses the smallest available memory size.
The request-response latency and cost of the three different kinds of invocations made during this experiment follow different distributions.
They show up as three steps in the cumulative distribution plots.
This shows that it might be useful to use different fusion setups per invocation depending on the root task, so that the setup can be changed depending on where a call comes from.
Overall, the optimization process reduced $rr_{med}$ from 59ms to 57ms, while cutting the average billed cost in half (from 1.9\md{} to 0.82\md{}).

\subsubsection*{\eid{Web}{Cold}}

\begin{figure*}
    \centering
    \includegraphics[width=\textwidth]{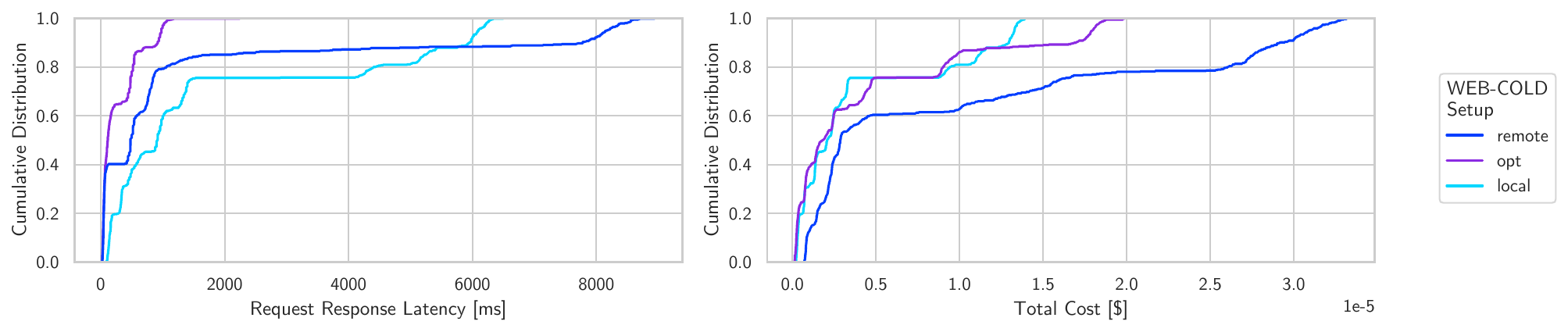}
    \caption{In the \eid{Web}{Cold} experiment, \sid{opt} has a lower $rr$ than \sid{remote}, and has a comparable cost to \sid{local}.}
    \label{fig:eval:shop:cold}
\end{figure*}

In the cold start experiment (cf.~\cref{fig:eval:shop:cold}), \sid{opt} has a lower $rr_{med}$ than \sid{remote} and \sid{local}.
Noticeably, \sid{local} has a similar cost distribution to the optimized setup, but has the highest $rr$.
The previously optimized setup also shows the best performance in the cold start experiment ($rr$ for \sid{local} = 2\md{}, \sid{remote} = 2.8\md{}, \sid{opt} = 1.8\md{}. $cost$ for \sid{local} = 918ms, \sid{remote} = 480ms, \sid{opt} = 113{}).

\subsubsection*{\eid{Web}{Scale}}

The experiment using a scaling workload shows a similar result to the previous experiment. \Sid{opt} has an $rr_{med}$ of 65ms, which is the same as \sid{remote}, and seven times faster than \sid{local} (460ms).
\Sid{opt} (0.7\md{}) is almost half as expensive as \sid{local} (1.3\md{}), which in turn is 69\% less expensive than \sid{remote} (2.2\md{}).

\begin{figure*}
    \centering
    \includegraphics[width=\textwidth]{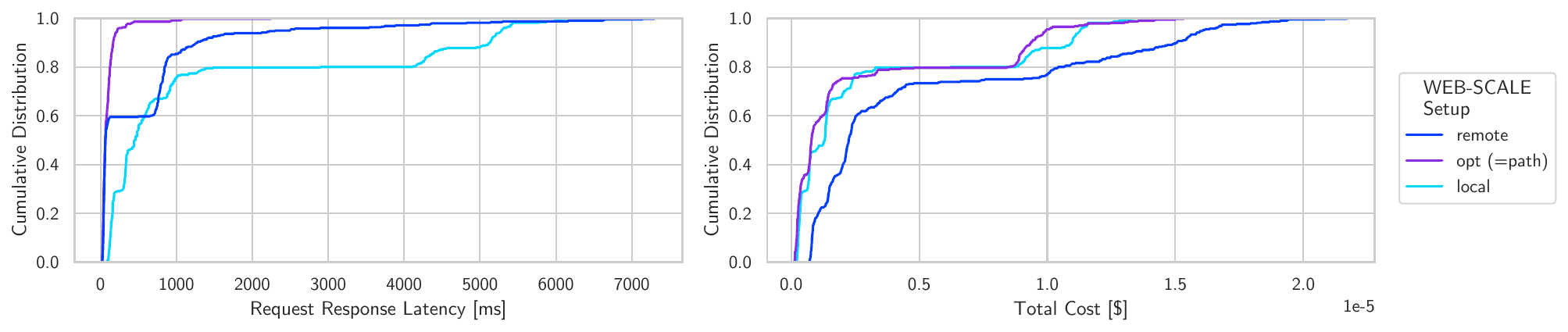}
    \caption{In the \eid{Web}{Scale} experiment, \sid{opt} has the lowest cost and fastest $rr$.
        While the total cost of \sid{local} is comparable to \sid{opt}, it has a worse performance.
    }
    \label{fig:eval:shop:bursty}
\end{figure*}

\subsubsection*{WEB: Discussion}

The results of the web shop experiments show that \name{} can be used to optimize complex real-world serverless applications, which have different user paths.
In this use case, the most cost-effective setup uses the smallest available memory size.
This increases the request-response latency compared to setups using bigger, more expensive infrastructure configurations.
In all three experiments, the setup found by the optimizer is significantly less expensive and faster than the other fusion setups.

\subsection{Framework Overhead}\label{subsec:overhead}

\name{} adds a handler to every function that manages calling the different tasks.
This adds an overhead for every function and task call.
In an experiment where we called a single empty task once per second for 200 seconds, the handler on average ran for 1.3ms when the function instance was already warm (standard deviation $\sigma=$ 1.24), and on average ran for 36.6ms in cold starts ($\sigma=$ 23.4).

While calling a task locally has almost no overhead, calling a task remotely requires additional time to send an HTTP request to another function, which takes $\leq50$ms.

The Optimizer adds no additional performance overhead to function calls, since it runs inside its own function and only reads the CloudWatch logs written by every function call.
Computing the next fusion setup for every 1,000 invocations takes around one second, while extracting the invocation data from CloudWatch sometimes takes considerably longer depending on the number of cold starts due to limitations in the CloudWatch API.
The CloudWatch limitations can be circumvented by adding platform support for the specific data extraction \name{} needs (requiring platform support), or by storing monitoring data in a serverless database (increasing cost).
Depending on the scale of the application, storing metadata about all calls might exhaust the resources of a function.
In this case, it is possible to sample metadata (reducing accuracy) or split computation into multiple functions each handling a smaller part of the input data.

The prototype also incurs additional cost, e.g., the Optimizer function and API Gateway costs, which we argue is not relevant for our evaluation since they are a consequence of our implementation choices and independent of architecture and conceptual approach:
Other FaaS providers might provide different methods to directly call their Functions that do not incur additional cost, while the number of requests and their relative execution duration is likely to stay consistent across different FaaS providers.

Although fusing several tasks into a fusion group increases deployment package size, our experiments have not shown this to inhibit cold start performance improvements.
First, this may be a result of the Optimizer only adding necessary files into a deployment package, whereas all dependencies (such as the Node.js runtime) are shared between tasks in a fusion group.
Second, Brooker et al.~\cite{Brooker_2023_AWSContainerLoad} have shown that increased deployment package size should not lead to proportionally increasing cold start latency on AWS Lambda.

\section{Discussion \& Future Work}
\label{sec:discussion}
In existing serverless cloud platforms, the task of sizing FaaS functions as well as picking the best resource amounts still needs to be handled by the developer.
In the \name{} framework, this is offered as an abstraction and is fully automated.
Even the comparatively simple heuristic, which we have implemented in our prototype -- remember that the optimization strategy was not the focus of our work -- could reduce cost and request-response latency in all experiments.
Nevertheless, some aspects of our approach warrant further research which we will discuss in the following.

\paragraph*{\textbf{Platform Integration}}

Our prototype is implemented on top of AWS Lambda and runs inside the FaaS functions.
This limits the information available to the framework.

While the information the Optimizer needs is easily accessible to the platform, exporting it takes additional time and overhead.
The same architecture could be implemented as part of the FaaS platform which could lead to increased performance due to additional information that would be available, e.g., for allocating tasks to functions or placing related function instances on the same physical machine.
Furthermore, call graph analysis could then be used to preemptively start functions in anticipation of tasks that will need it to avoid cold starts.

Intricacies of the platform also influence the Optimizer algorithm.
For example, AWS Lambda scales computing power linearly with configured memory.
In contrast, Google Cloud Functions increments computing power in steps of full vCPUs at certain memory sizes.
Here, step-wise CPU scaling would make some resource options significantly more cost-efficient.
Different platforms thus require specialized resource optimization approaches.
Additionally, the FaaS platform knows the current resource utilization of the underlying machines which could be used to further optimize the fusion setup.
While integration into the FaaS platform could increase performance, our approach also works when deployed by application developers and can be used until function fusion is supported by platforms.

\paragraph*{\textbf{Hardware Acceleration}}
A further avenue of research is the addition of hardware accelerators to serverless architectures~\cite{Werner_2022_Hardless,Pemberton_2022_KaaS}.
Some FaaS platforms offer optional support for hardware accelerators such as GPUs that can be used by functions.
While some tasks can run significantly faster as part of FaaS functions with access to hardware accelerators, they are also more costly to run and lead to increased cold start times.
\name{} could be extended to handle the grouping of tasks to functions with or without hardware accelerators to further optimize applications.

\paragraph*{\textbf{Programming Model}}

In previous work~\cite{Scheuner_2019}, we have presented an approach using transpilation to fuse tasks into fusion groups, rather than the fusion handler we use in our prototype.
In both approaches, the code points that are suitable for function fusion, i.e., task entry points, need to be clearly marked by developers.
Thus, they are mainly intended to be used when developing new applications and not to transform legacy applications into serverless applications.
Spillner~\cite{Spillner_2017} presented a framework that transforms a Python application into (FaaS) functions which could be a preprocessing stage for \name{}.

Another avenue of research is support for polyglot applications, i.e., applications written in more than one programming language.
Currently, our prototype of \name{} (but not the approach itself) assumes that all tasks are written in the same language as the fusion handler.
Using WebAssembly~\cite{Marcelino_2023_Cwasi} or other runtimes such as Docker, it is possible to execute polyglot applications in one runtime, which makes function fusion possible on platforms that support these runtimes.
This would require adding fusion support to these runtimes and writing a shim that communicates with the runtime for every programming language that tasks can be written in.
\paragraph*{\textbf{Fusion Groups \& Infrastructure Optimization}}

In our approach, fusion setups are determined only by information about previous invocations, leading to a performance profile of the application.
These fusion setups are static in the sense that they only change after Optimizer iterations.
Yet it may also be feasible to select a fusion setup based on the type of invocation.
For example, the fusion handler could change its behavior when it detects a cold start or when the input data matches certain properties.
Our experiments show that the all local setups -- if possible resource-wise -- can be less expensive than the optimized fusion setup during cold-start heavy workloads.
In the current implementation, the Optimizer uses latency and memory consumption of every invocation as input.
Future implementations of the Optimizer could take more parameters into account and use sampling to reduce the input size for high-scale applications.
An alternative could be training a machine learning based on data from multiple fusionized applications to let the Optimizer recommend a good strategy directly in the first step before exploring whether it is indeed a local optimum.
Nevertheless, any additional dynamic behavior will add further complexity and overhead to the fusion handler which will at least partly offset the benefits that can be achieved.
Application updates are in the current prototype handled by re-setting the optimization state.
Future versions of the Optimizer could, however, also use measurement data from previous versions of the source code as input parameter.
This would require analyzing the source code for changes to invalidate the tasks that have been changed.

\paragraph*{\textbf{Experiments}}\label{subsec:discussion:eval}

In our experiments, we have shown that \name{} can optimize applications in real-world scenarios.
For this, we used CPU-bound mathematical operations to stress the CPU.
This makes it easier to compare different levels of CPU usage and performance does not depend on external services such as object storage which could influence latency.
We analyzed how these services impact \name{} by using DynamoDB to store data in the IoT and web shop experiments.
Some applications, however, might also be influenced by big changes in the performance of other components, which we did not check in our experiments.

Overall, we believe that using \name{} leads to significant performance and cost improvements for the majority of applications.
For all other applications, the monitoring results of \name{} could be used to fall back to the baseline in which all tasks are deployed as their own task.
This way, it would be guaranteed that using \name{} never leads to worse performance and cost.

\section{Related Work}
\label{sec:relwork}
Scheuner and Leitner~\cite{Scheuner_2019} proposed the concept of function fusion.
In their vision, application code is automatically broken up into functions, which are then deployed on a serverless platform.
This works with existing applications, as developers do not need to add special markers where function fusion could happen.
During our development, we found that this approach has practical limitations: Developers might add indirect data dependencies to their code, e.g., by using public variables of other source code modules.
To support this kind of data access, \name{} would need to transfer more state between functions, which adds significant overhead.
Thus, we decided to make these dependencies explicit by allowing developers specify task boundaries.

Elgamal et al.~\cite{Elgamal_2018} present an algorithm that minimizes the cost of functions while keeping latency below a fixed threshold by using function fusion as well as placing the functions at specific edge locations using AWS Greengrass.
Their approach uses AWS Step Functions, which they identify as major cost driver in their implementation.
With WiseFuse, Mahgoub et al.~\cite{Mahgoub_2022_Wisefuse} present a similar vision of function fusion: they propose to co-locate parallel function instances with data dependencies on the same server to minimize communication overhead, and fuse subsequent tasks in the same function.
In their usage model, users specify their (unoptimized) DAG, with which either the cloud provider runs profiling runs free of charge and then suggests an optimized DAG to the user or users run this optimization themselves.
In comparison, \name{} works without user interaction to iteratively improve the application and does not need any test-runs as live monitoring data is used.
The approaches by Elgamal et al.~\cite{Elgamal_2018} and Mahgoub et al.~\cite{Mahgoub_2022_Wisefuse} both require that workflows can be modeled as a directed acyclic graph (DAG).
This means that the architecture of the application is limited as tasks calls cannot go in both directions, not even transitively.
Additionally, developers are usually expected to specify the DAG alongside their application code and are responsible for keeping them up to date.
In comparison, \name{} can work with any shape of call graph and automatically identifies it based on monitoring data, reducing efforts for developers.

Other works optimize serverless workflows without requiring a DAG.
Burckhardt et al.~\cite{Burckhardt_2022_Netherite} use stateful functions to implement workflows that have support for execution guarantees and concurrency control.
Daw et al.~\cite{daw2020xanadu} minimize cascading cold starts in workflows by pre-warming functions.
As in \name{}, they also can extract the call graph from monitoring data.
Since they optimize different parts of the application, all three approaches are complementary and can be used in parallel. 

While we target applications composed of multiple tasks in this paper, reducing latency and cost of executing a single serverless function without considering composition knowledge, e.g., by reducing cold starts, has been discussed in multiple previous studies~\cite{Manner_2018_Coldstarts, Bardsley_2018_coldstarts, paper_bermbach_faas_coldstarts}.
Others have used statistical analysis~\cite{Akhtar_2020}, machine learning~\cite{Eismann_2021_Sizeless,Moghimi_2023_Parrotfish}, or profiling~\cite{Cordingly_2022_memorysizes} to predict the optimal infrastructure configuration of a single function by only looking at some function configurations.
This reduces the number of experiments for determining the optimal infrastructure configuration and could be used in conjunction with \name{} to further speed up finding an optimized setup.
The CPU and memory footprint of functions can be reduced by sharing memory between functions running on the same virtual machine~\cite{Mahgoub_2021_SONIC}, by using application-level sandboxing~\cite{Akkus_2018_SAND}, or by using unikernels~\cite{Fingler_2019_Unikernel}.
Mahgoub et al.~\cite{Mahgoub_2022_Orion} additionally reduce the resource consumption of function invocations by placing multiple parallel invocations inside one function instance.
These ideas are complementary to \name{} and may further improve performance of deployed fusion groups.

In this work, we assume that tasks are supposed to be deployed as serverless cloud functions, yet previous work has also considered the optimal placement of tasks over a varied set of compute services such as Container-as-a-Service platforms~\cite{Czentye_2019}, virtual machines~\cite{Horovitz_2019_VmMlFaaS}, across hybrid clouds~\cite{Khochare_2023_XFaaS}, or in a fog environment~\cite{Pfandzelter_2019_FogProcessing,paper_pfandzelter_zero2fog}.

Ali et al.~\cite{Ali_2020_Batching} propose a method for optimizing cost and minimizing latency by batching multiple invocations into a single function execution.
Such an approach is especially useful if cold starts account for a significant part of the application duration, e.g., if the function needs to download big datasets during startup.
While \name{} focuses on optimizing single requests, it could be combined with the work of Ali et al. in cases where increasing the latency of some invocations is acceptable.
The general approach of giving the platform access to more application knowledge, enabling the platform to change it own and the applications' behavior, has been used in serverless applications to delay the execution of functions during times of high load~\cite{schirmer_2023_ProFaaStinate,Sahraei_2023_XFaaS}, and to optimize cloud virtual machines for their specific workload~\cite{Huang_2024_WorkloadIntelligence}.

\section{Conclusion}
\label{sec:conclusion}

In this paper, we have presented the \name{} approach and our proof-of-concept implementation.
\name{} is an approach for removing operational burden from developers by automating the process of turning the tasks written by developers into an optimized FaaS application.
Calls from one task to another task can be fused and executed inside the same function (reducing call overhead and mitigating cascading cold starts) or can be handed off to another function (improving how resources can be allocated).
Leveraging monitoring data, \name{} optimizes the distribution of tasks to functions as well as the infrastructure configuration to incrementally optimize deployment goals such as request-response latency and cost.

Further, we present a heuristic for the iterative optimization of the FaaS application.
Our heuristic first optimizes the path invocations take by fusing tasks together, and then optimizes the infrastructure configuration of the resulting functions.
Using a proof-of-concept prototype of \name{} for the Node.js runtime on AWS Lambda, we have shown that our heuristic can improve request-response latency as well as cost by more than 35\% in real-world use cases.
In future work, we plan to develop further Optimizer strategies, both using this prototype and integrated into open source FaaS platforms.

\section*{Acknowledgment}

We thank Jun-Zhe Lai who supported this work in the scope of a master's thesis.

\bibliographystyle{IEEEtran}
\bibliography{bibliography}

\begin{IEEEbiography}[{\includegraphics[width=1in,height=1.25in,clip,keepaspectratio]{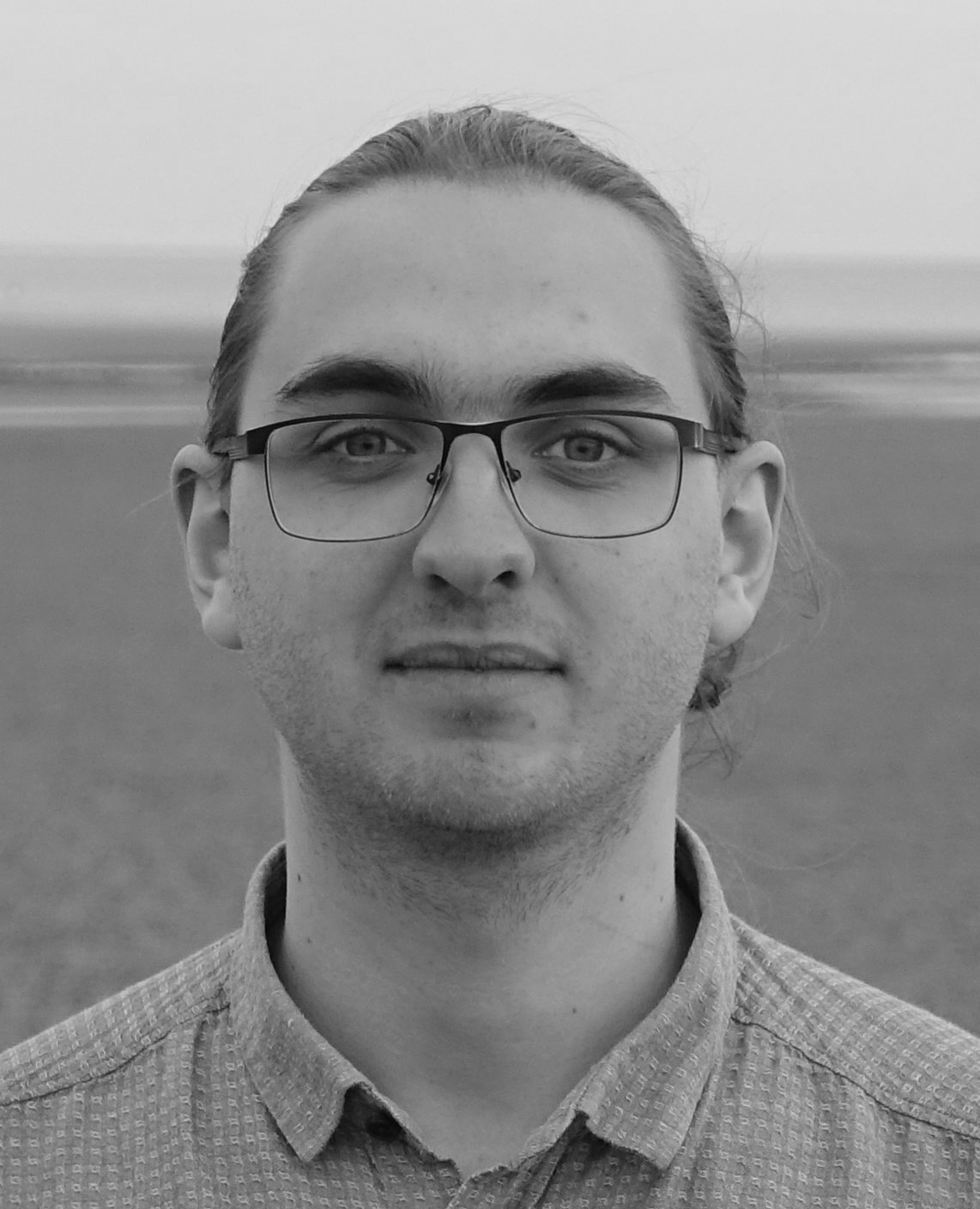}}]
    {Trever Schirmer}
    received the master's degree in information systems management from TU Berlin, where he is currently working toward his PhD degree with the Scalable Software Systems research group. His research focuses on optimizing serverless applications and platforms. Before starting with his PhD, he worked as a software developer.
\end{IEEEbiography}

\begin{IEEEbiography}[{\includegraphics[width=1in,height=1.25in,clip,keepaspectratio]{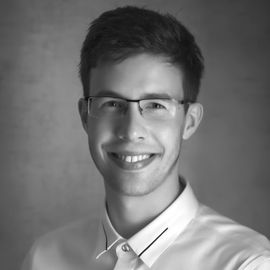}}]{Joel Scheuner}
    is a Software Engineer at LocalStack. He previously received his PhD Degree at the Internet Computing and Emerging Technologies Lab (ICET-lab) at Chalmers University of Technology, Gothenburg, Sweden.
\end{IEEEbiography}

\begin{IEEEbiography}[{\includegraphics[width=1in,height=1.25in,clip,keepaspectratio]{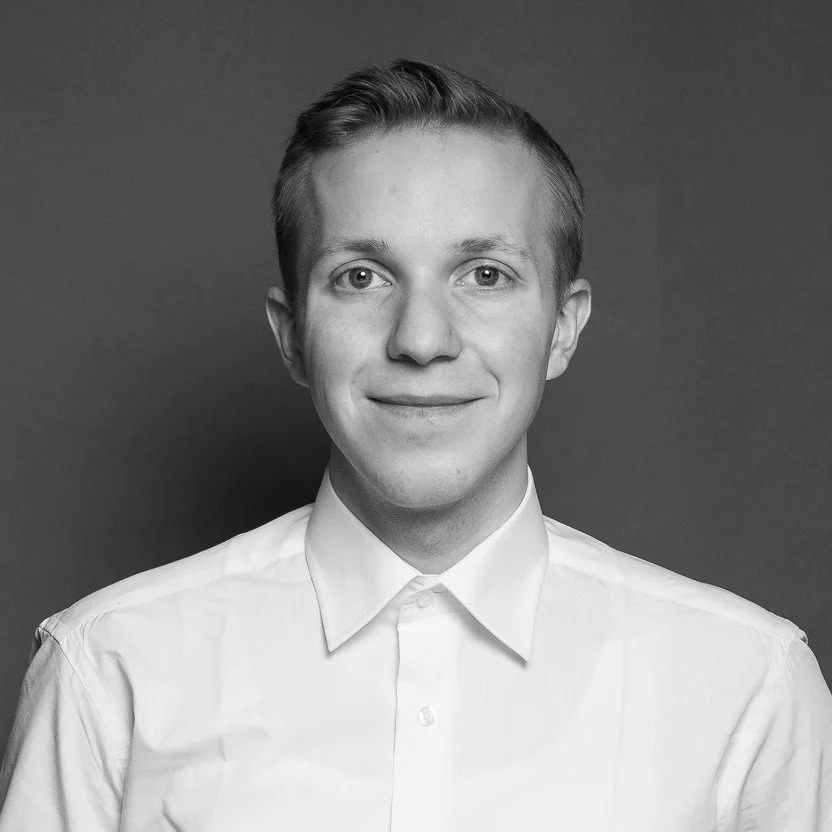}}]{Tobias Pfandzelter}
    has been a research associate and PhD student at the Scalable Software Systems research group since September 2019. Before that, he completed his Bachelor and Master in Computer Science at TU Berlin.
    His research focus is on edge computing in LEO satellite constellations.
\end{IEEEbiography}

\begin{IEEEbiography}[{\includegraphics[width=1in,height=1.25in,clip,keepaspectratio]{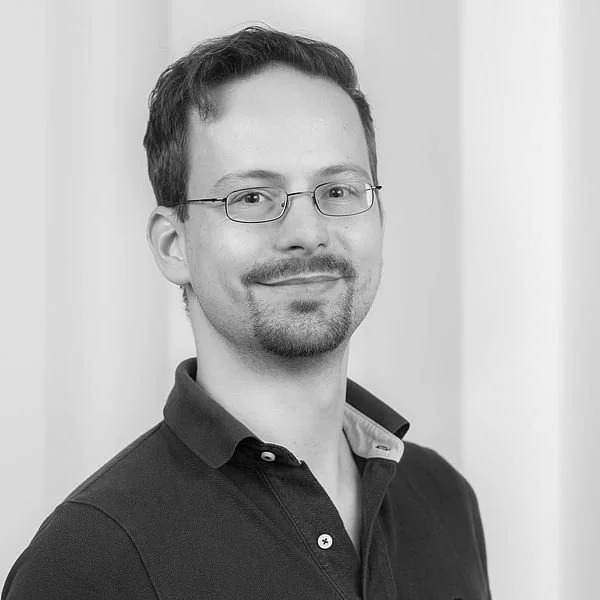}}]{David Bermbach}
    received a diploma in business engineering and a PhD degree, with distinction, in computer science from Karlsruhe Institute of Technology.
    He is currently a full professor at TU Berlin as well as the Einstein Center Digital Future and is heading the Scalable Software Systems research group.
    His research focuses on benchmarking and platforms and applications for cloud, edge, and fog computing.
\end{IEEEbiography}

\balance
\end{document}